# Real Time Enhanced Random Sampling of Online Social Networks

Giannis Haralabopoulos, Ioannis Anagnostopoulos

University of Central Greece, Dept. of Bioinformatics, Lamia, Greece

ghara@ucg.gr, janag@ucg.gr

**Abstract.** Social graphs can be easily extracted from Online Social Networks (OSNs). However, as the size and evolution of this kind of networks increases over time, typical sampling methods used to evaluate large graph information cannot accurately extract network's properties. Furthermore in an attempt to deal with ever increasing access and possible malicious incidents (e.g. Denial of Services), OSNs pose access limitations in their data, making the crawling/sampling process even harder. A novel approach on Random Sampling is proposed, considering both limitation set from OSNs and resources available. We evaluate this proposal with 4 different settings on 5 different test graphs, crawled directly from Twitter. Results show that every scenario needs a different approach. Typical Random Node Sampling is better used for small sampling sizes, while Enhanced Random Node Sampling provides quicker and better results in larger samples. Still many questions arise from this work that can be considered as future research topics.

## 1 Introduction

Online Social Networks (OSNs) provide a great source for social graph analysis, but due to their dynamic nature, advanced methods and lots of resources are required for processing their statistical properties. This dynamic capacity is demonstrated in Twitter, which climbed 22 spots in the world's most visited sites within a year, it currently is the second most visited OSN [10][11]. As such, every attempt to estimate its graph properties is getting even harder.

Various sampling methods used throughout the years are not able to accurately describe a social graph's structure and properties. These methods analyse and evaluate subgraphs, which only consists of the sampled nodes. For example, although random sampling is widely popular in graph sampling processes, it fails to accurately project most graph properties [9].

In addition, Twitter – and various other OSNs – imposes an hourly limitation of 350 requests per hour (for authenticated users), thus making data acquisition a costly procedure, resource wise. Despite that, if we effectively use these requests, various sampling methods can still be efficient and further improve our sampling results.

So, the issues that must be addressed are described in the following questions:
- How can we fully utilise our limited graph crawling requests?
- Can sampling methods provide sufficient data about an OSN graph?
- Can we enhance existing sampling methods to further improve results?

The answer to all three questions is a real time enhanced sampling process. That is, a selective process that takes request limiting into consideration coupled with enhancements to improve our results.

This work aims to test two enhancements on the random sampling method. Our proposal makes use of all the information available in a single node of an OSN. In addition, it is executed in real time and in parallel with the crawling process, thus saving valuable resources. Lastly, the proposed enhancements can be used with any graph sampling method.

The OSN on which we tested our proposal is Twitter. Within a 25-day time period we crawled data (called from this point herein as Test Graphs - TGs). To be more specific the crawling process started on $5^{th}$ of March 2012 and ended in the $30^{th}$ of the same month. The method we used was Breadth First Crawling, with a minimum number of nodes to be fully analysed of 25K.. They were randomly chosen and our calculations showed that we fully analysed up to the fourth degree neighbours for each seed. A user in Twitter corresponds to a graph's node and the user's followers/following to in/out edges of that node.

The first enhancement we propose is based on the fact [12] that social networks tend to follow a Pareto Distribution [7]. We tested three different degree distributions, which could improve sampling results. Since we do not have any knowledge of graph properties we are investigating the area surrounding Pareto 80 – 20 distribution. . We prove that our implementation is not bound to any a-priori known properties, since it is applied on the fly along with the crawling process.

The second improvement is based on the observation that when we sample a graph and we only use the resulting subgraph for analysis, we lose a vital information considering the hard request limitations we face. Thus, we propose to add all the neighbouring nodes of a sampled node to the subgraph, which in fact is the combination of two techniques, random and neighbourhood sampling.

In our work, the evaluated graph properties are the Number of Edges, Mean Degree, Clustering Coefficient, Assortativity, Number of Components and time required for the process. We compare the properties of each TG with sampled mean value of multiple iterations, and in order to get a better idea on time handling, we provide another comparison of results along with time required to produce them.

## 2   Related Work

Graph analysis has been the subject of many essays, while it is a significant topic in many real-life applications in both technological and social field. From simple metrics to advanced trend prediction, graphs can provide valuable information through OSN analysis. However, the amount of disseminated information is vast, thus suggesting graph sampling an important process for analysing OSNs structure and properties.

Leskovec and Faloutsos in [1] provided an extensive analysis and evaluation of graph sampling. Their focus is graph patterns along with graph properties. However graphs are relatively small (3 to 75 thousands nodes) but probably more cohesive than our TGs. Ahn et al. in [2], analyse huge graph's structure based on graph properties and degree distributions. Having access to the full graph of an OSN provides valuable information and a great benchmark for the sampling methods. Furthermore, they compare results and properties of their full OSN with 3 other OSNs of much smaller size.

Dasgupta et al. in [3], deal with community identification of huge information networks. Many different databases are analysed with the graph conductance playing a key role in conclusions. Huberman et al. in [4] deal mostly with Twitter properties, rather than actual graph properties.

Noordhuis et al. provided an insight of mining twitter and further executing PageRank to their dataset [5]. Another important work in respect to deep graph analysis is authored by Broder et al [6]. Although this research conducted more than one decade ago, it's still relevant to modern datasets and networks. Lastly, Ye et al. [8] analyse the crawling process of a graph, mainly on the part of edge and node discovery. The authors also address the issues of multiple seed choices and protected users, which affect both the crawling and sampling process of an OSN.

In this research, graph theory and sampling are combined with web crawling. The practical problems of crawling Twitter and the evaluation of its sampling, should be analysed in detail with the focus on efficiency rather than results. Various methods and analysis schemes used in this work are inspired by the aforementioned related work, and they are referred accordingly upon their application in the following sections.

## 3 Methodology

### 3.1 Dataset and Resources Description

In this work, we have chosen Twitter as the OSN in test, not only because it is constantly evolving scale and property wise, but also because it has a strong social impact [13][14]. It uses directed link information and from technical point-of-view it provides an open API for data acquisition (despite some limitations discussed in previous section). Twitter users can either follow someone or be followed by someone, identically to a node's out and in edges. The analysis of user linking associations provides the in and out degrees in respect to the connected users. Since Twitter does not provide further information about its actual graph properties, research teams and communities have to crawl and analyse their own data for evaluating their methods.

Thus, during a three-week period (between March 5th and March 30th of 2012), we fully analysed virtually 200K Twitter users in order to discover their links. Upon concluding the crawling process, approximately 60GBs of data were obtained containing more than 93M discovered users and 570M edges.

For the crawling process we used 30 different starting seeds (30 Greek users with the highest number of followers) and then performed a breadth first sampling. For every starting seed, we fully analysed its 2$^{nd}$ level neighbours, while for some seeds we analysed their 3$^{rd}$ and 4$^{th}$ level neighbours. Since our goal was to test our algorithms on several datasets, we chose not to analyse the dataset as a whole, but instead we kept data separated based on seed nodes.

Data was kept separated, based on the seeding node. This allowed us to analyse each part fairly quickly, using R project [15] and igraph package [16]. We compiled 5 different sampling methods and 6 graph property extraction and visualization algorithms in R format. Subsequently we used igraph package to test them, which is very fast but requires heavy utilisation of RAM. We ended up using 18 to 36GB of RAM in order to analyse each TG. Due to RAM requirements and multiple algorithms iterations needed, we decided to use Amazon Web Services high memory instances so we could parallelize the analysis.

### 3.2 Proposed algorithmic enhancements

Both our enhancements can be applied to any sampling process, however in this work we have tested them only according to random sampling method, since it is commonly used, quite fast and provides reliable and robust results. The main drawback of this method, is its inability to maintain the power law distribution of a graph $G$.

Sampling a graph $G$ provides some nodes and some edges. Usually after sampling, the next step is to find the subgraph consisting only of $N$ *nodes* -with $E$ *edges* amongst them- and then analyse the resulting graph. The problem of this subgraph approach is that it discards many discovered edges $E'$ along with their nodes However, in the followings we will show that both $N'$ and $E'$ can improve the accuracy of random sampling method.

Initially, upon random sampling on $G$, we test if we can use all discovered nodes and edges to our advantage. So, adding a node to the subgraph along with all its adjacent edges and neighbouring nodes, is the first proposed enhancement.

A power law distribution of a graph can be estimated by various methods such as Maximum Likelihood Estimation in [17] and Kolmogorov–Smirnov estimation in [18], but almost every estimation method requires full -or at least some- knowledge of the graph and its properties. Unfortunately, upon crawling Twitter graph, the only known property is the in/out degree of each node. Therefore, most of these methods cannot be applied for sampling purposes.

The second enhancement we propose, depends on whether by accepting/rejecting sets S of nodes to the subgraph (based on their degree distribution), will lead in further sampling process improvement, where $S$ has a predefined number of nodes to be used in every set. In this sampling proposal, we check if the top-$A$ percentage of a node set has the $B$ percentage of degree's sum. $B$ and $A$ are coupled in 3 different distributions, namely 85–15, 80–20, and 75–25. This means that in 85-15 distribution, nodes are accepted into the sampling pool only if the top-15% of highest degree nodes has the 85% of the set's total degree, while similarly in case of 80-20 or 75-25 distributions, nodes are accepted if the-top 20% or top-25% of highest degree nodes has the 80% or 75% of the set's total degree respectively.

Both of our improvements are proposed in an effort to further enhance sampling results and overcoming the limits imposed by OSNs. These limits make each crawled node valuable and the requests are to be considered as a limited - but necessary - resource. To be specific, in Twitter one request is used to:

a) Obtain the number of followers and following of a user, thus essentially discover a node's in and out degree, and

b) Obtain linked (followers or following) users in sets of 5K, discover a node's in or out neighbours 5.000 at a time.

This means that in order to get all followers of a user with 10.000 followers and 5.000 following, we would use four requests; one to obtain the node's in/out degree and three to obtain its list of adjacent nodes..

### 3.3 Real Time Enhanced Random Sampling

In our research, sampling and crawling can be considered as a whole, since the way we acquire data defines our sampling method. The sampling method we test apply is separated in two-flow sub-charts. The enhancements are drawn with dashed lines, while the typical crawling/sampling process is illustrated with solid lines (Fig.1).

The first sub-chart is the usual crawling process of an OSN and consists of three separate steps:

**1.** Crawl a node
**2.** Add node to selection database
**3.** Methodically pick next user. We should note here, that step 3 differentiates upon the method used (e.g. for Breadth First crawling we would use the first crawled neighbour, while for Random crawling we would randomly choose one).

The second sub-chart consists of our proposed enhancement steps along with the typical conditional symbol for sampling threshold control. Each step will be reviewed in detail.

**4.** Using a predefined number of nodes, we choose a random set $S$ from the total crawled nodes. In our tests, we chose to use $S=2\%$ of the sample's size we need, since this percentage is ideal for frequent monitoring of the distribution check and adds nodes to the sampled pool (step 7) relatively quick.

**5.** At this point, we check if the set of nodes follows the desired distribution, based on the in and out degree this node has on the original OSN. If the set follows the desired distribution, then it is added to sampled pool (step 7*)*, while if not, we move to loop control (step 6). At the end, we perform tests for the three different distributions, starting with 85–15, then 80–20 and lastly 75–25.

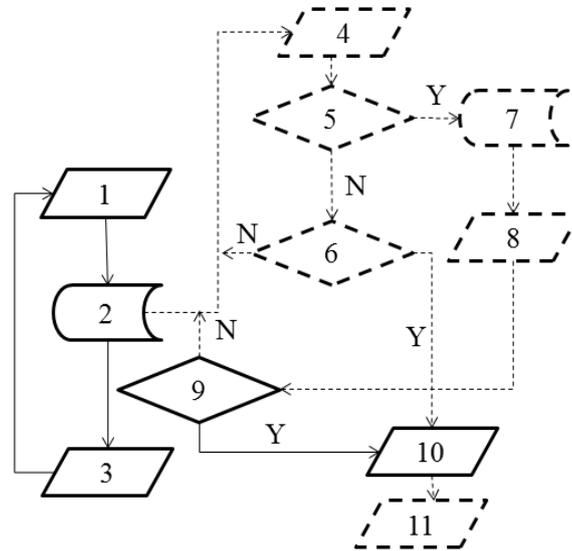

**Fig. 1.** Algorithm's Flow Chart

**6.** Another conditional junction that prevents the node set distribution check (step 5) from looping. So, if our predefined distributions do not exist in the crawled graph, the algorithm stops, using the nodes already added to step 7 in order to create the sampled graph. The number of tries to discover a distribution is $n=10$ and it is controlled through step 8. For example, in case that we find a desired distribution in the $9^{th}$ try, we will then have 10 more tries to find another set that satisfies our distribution rules.

**7.** This is the sampled pool. The nodes added to this pool will be used to create a Sampled Graph.

**8.** The point we reset the number of tries remaining, as described in step 6.

**9.** This conditional junction tests if we reached the sampled threshold. During our tests, we used various different sample sizes to evaluate our method, ranging from 10% up to 90%, with a 10% step.

**10.** Here the subgraph is created from nodes in the sampled pool. This is the last step of Random node sampling (RNS).

**11.** Adjacent nodes and edges -to every node in the subgraph- are added to the sampled graph. The adjacent nodes and edges can be discovered via the respective requests to the OSN. This is the step that defines Random node sampling all edges (RNSE).

All of the above steps are independent of the sampling/crawling method, meaning that this algorithmic process can be applied to every sampling method, without any structural changes.

## 4 EXPERIMENTATION

### 4.1 Testing

Fourteen graphs were tested in 8 variations of our proposed algorithm. These are the following:
1. RNS
2. RNSE as described in Section 3.3.11
3. RNS with 85–15 selective distribution (RNS 85-15), see Section 3.3.5
4. RNSE 85–15
5. RNS 80–20
6. RNSE 80–20
7. RNS 75–25
8. RNSE 75–25

Each variation was tested 10 times by 9 different sampled percentages. Here, we present only the mean relative error, but we have to note that variance was very small (<0.01%) for every iterations on the same TG. The algorithm was applied 10800 times and the calculation time was approximately 800 hours. When compared to relevant work [8], our approach seems quite fast but requires large memory pools, as mentioned above.

Table 1. Test Graph Values

|  | Edges | Vertices | Mean Degree | Mean Clustering Coefficient | Mean Assortativity |
|---|---|---|---|---|---|
| Max | 69872473 | 8444642 | 19,61 | 0,022451 | -0,31064 |
| Average | 40731274 | 6643998 | 12,14 | 0,013787 | -0,44001 |
| Min | 25658132 | 4662987 | 9,74 | 0,007862 | -0,51234 |

Every graph part we obtained was tested, their mean values are presented along with min and max values to provide a measure of graph scale (Table 1). These properties and their values show that all graphs have a dissortative pattern and are not densely connected. Furthermore, we have to note that every graph has only one component, which is a direct result of the crawling method we used (Breadth First).

On every error chart (Figures 2, 4, 6 and 8 in the Appendix), we distinguish the results according to the variables taken into account. Essentially, two different sampling techniques are used with four different evaluations. In one hand, RNS and RNSE consider the mean degree, the clustering coefficient and the assortativity. On the other hand, RNS ED and RNSE ED consider the previous variables with the addition of edge discovery relative error. This presentation was selected in order to highlight the benefits of our proposed sampling scheme, mainly in cases where edge discovery is as crucial as every other property of a sampled graph.

In respect to each error chart, we provide a "result improvement over time" chart where the percentage improvement scaling is illustrated accordingly. It is obvious that in every tested case, increasing the sample size has exclusively positive results.

### 4.2 Results – Discussion

On every test, we noticed that RNS (either with or without selective distribution) sustained the clustering coefficient and assortativity values throughout the different sample sizes, while it failed completely on mean degree and component size. Since our initial graphs have already low clustering coefficient and assortativity, we are unsure on whether this was related with our data. However, while on one hand RNSE maintained a fairly stable mean degree and component size for different sample sizes, on the other hand the clustering coefficient and assortativity were marginally improved on sample sizes greater than 20%. Furthermore, edge discovery evaluation is heavily favoured by RNSE, with a mean accuracy improvement over RNS of 40%.

**Table 2.** Mean percent error for non-selective sampling

| Method / Sample size | RNS | RNSE | RNS ED | RNSE ED |
|---|---|---|---|---|
| 10% | **29,85%** | 41,93% | **46,91%** | 49,96% |
| 20% | **25,45%** | 29,48% | 42,41% | **35,84%** |
| 30% | 21,81% | **21,21%** | 38,08% | **26,07%** |
| 40% | 18,77% | **15,19%** | 34,00% | **18,93%** |
| 50% | 16,28% | **11,07%** | 30,26% | **13,89%** |
| 60% | 10,46% | **4,35%** | 20,64% | **5,54%** |
| 70% | 8,51% | **2,90%** | 17,18% | **3,70%** |
| 80% | 4,81% | **0,86%** | 9,93% | **1,11%** |
| 90% | 2,37% | **0,20%** | 5,02% | **0,26%** |

Regarding the results from non-selective distribution random sampling (Table 2), we can distinguish the results based on our sampling goals. When our aim is, to accurately sample a social graph based only on its properties (mean degree, assortativity and transitivity), RNSE is more accurate for sampling sizes greater than 20%. Similarly, when we have to consider edge discovery as well as graph's properties, then RNS provides better results only for sample sizes up to 10%. Furthermore RNSE ED accuracy over RNS ED is growing exponentially as the sample size increases. The same results are demonstrated on the best selective distribution case, 80 – 20 (Table 3). Although the improvement of results for RNSE and RNSE ED start from 30% and 20% sampling sizes, for the respective cases. The loss of accuracy on RNS ED and RNS ED evaluations, is due to the high percent error of edge discovery compared to the mean low percent error of the other 3 properties. This connection is maintained throughout the different sampling sizes, but on sizes greater than 70% the effect is fading and the results of ED are almost identical to those of non-ED evaluations.

**Table 3.** Mean percent error for 80/20 selective sampling

| Method<br>Sample size | RNS<br>80/20 | RNSE<br>80/20 | RNS ED<br>80/20 | RNSE ED<br>80/20 |
|---|---|---|---|---|
| 10% | **31,48%** | 46,80% | **48,36%** | 55,34% |
| 20% | **27,52%** | 35,07% | **44,63%** | 42,27% |
| 30% | **24,12%** | 25,92% | 40,83% | **31,67%** |
| 40% | 20,62% | **18,50%** | 36,45% | **22,85%** |
| 50% | 17,11% | **12,46%** | 31,56% | **15,58%** |
| 60% | 13,70% | **7,76%** | 26,23% | **9,79%** |
| 70% | 10,19% | **4,20%** | 20,33% | **5,37%** |
| 80% | 6,88% | **1,79%** | 14,05% | **2,32%** |
| 90% | 3,45% | **0,38%** | 7,09% | **0,51%** |

On all three selective implementations, we used fewer vertices to create the sample graph (Table 4) when compared to those used in non-selective RNS and RNSE. Thus, we should consider the results in relation to the "economy" of recourses. As mentioned previously, in Twitter API one request is needed for the analysis of a node's degree (in and out) or for the discovery of 5K neighbours (in or out). In the worst case scenario, for each sampled size, we managed to use 111 less requests, while in the best case, we preserved half a million requests. Even in a moderate case where 10K less requests are used, 27 less hours are needed to discover the same part of the OSN and sample it. We can only imagine the scalability of resource conservation when this method is used for larger datasets. But, what is the cost of economy when results are needed?

**Table 4.** Mean vertices conservation

| Method | Mean Vertices Used | Mean Percent Error |
|---|---|---|
| 85-15 | 3387917 | 51,69% |
| 80-20 | 4072877 | 16,16% |
| 75-25 | 4071613 | 16,59% |
| Non-Selective | 4403086 | 14,75% |

To answer this question and accurately define that cost, we will have to go through every different distribution we tested. First of all, in the case of 85–15 we needed 23.06% less vertices to get 51.69% less accurate results. For 80-20 and 75-25 we used 7.5% less vertices to get 16.16% and 16.59% less accurate results respectively. However the greater the sample size, the lower the loss of accuracy. The reduction in vertices used results in less time dedicated to crawling/sampling, with a moderate time gain of 25 hours per sample analysed, which equals to 300 hours for all of our 14 TGs. Furthermore in time/results charts we can see that RNSE scales results in relation with time much better than RNS on all four occasions. Time wise, all selective methods are slower than RNS. Similarly RNSE is slower than RNS on a smaller scale. To conclude, it is apparent that RNSE outperforms RNS in every case, either with selective distribution or not.

## 5   Conclusions – Future work

New limitations force us to reconsider our algorithms and find new ways to enhance our methods. We demonstrated a novel approach to crawling/sampling an OSN. Although not applicable to every research scheme, it provides sufficient –if not better- results. Sampling can be combined with crawling in a way that we conserve our time and computational resources.

As far our research showed us, we would not propose one method for every sampling scheme. For crawling an OSN with small sampling size (<20%), we estimate that RNS is the most efficient sampling method. If we aim to crawl an OSN that does not have request limitations and we seek larger sample sizes (>20%) the RNSE method fits perfectly. While on most occasions where OSNs impose requests limitations and we are unsure of the sample size, we believe that the best scheme to use is the RNSE 80-20, since we not only conserve resources (requests and time in our case), but also we are not losing in terms of accuracy. Furthermore sampling results of RNSE 80-20 on samples greater than 70% where almost identical to the most accurate method (RNSE).

Still many questions arise from this initial research. Can we implement this method in crawling "undirected" OSNs? Will this proposal work on denser and/or assortative graphs? What about graphs with more components? Can we use different distributions? Can these enhancements work effectively on different sampling methods? Unfortunately, our resources are not unlimited.

All these questions are research subjects for the future. We will work towards analysing multiple OSNs, one at a time. Our direct goal is to establish whether the results were network dependant or this random sampling method will work on every OSN.

Twitter is evolving on an hourly basis, creating a vast social graph with so much information. Unfortunately this expansion enforces new policies from Twitter but most of these restrict access to Twitter's data, making is the only way to overcome these restrictions. Likewise, sampling is a solid base for trend analysis, a procedure which gets harder every day. Using proposed sampling schemes for analyzing and discovering user's links is not a necessary condition, but surely is a sufficient one.

# Appendix

On relative error charts we can see a comparison of the mean percentage error on 4 different occasions. RNS and RNSE label lines, show the mean percentage error of three properties -Mean Degree, Clustering Coefficient and Assortativity. Similarly, RNS ED and RNSE ED label lines, present the mean percentage error of Mean Degree, Clustering Coefficient, Assortativity and Edges Discovery.

On the variation charts, we can see the percentage variation of time, having as starting point the time required for the sampling and analysis of 10% sample size. Result plots have as a starting point the accuracy of the results derived from analysing samples of 10% size.

## A1. Without Selective Distribution

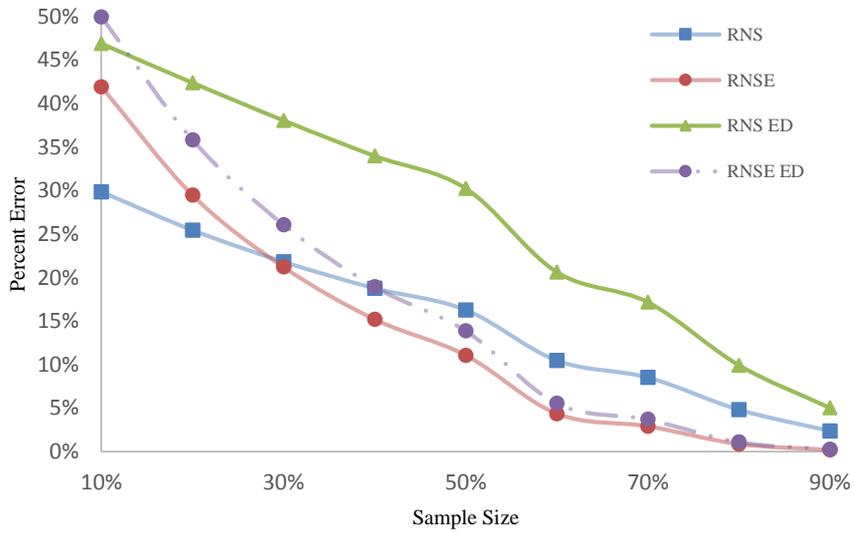

**Fig. 2.** RNS and RNSE Relative Error

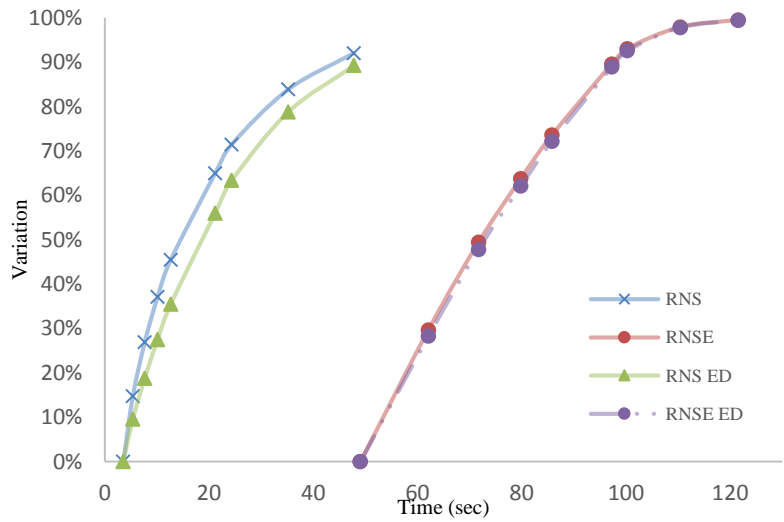

**Fig. 3.** RNS and RNSE Result Improvement over Time

## A2. 85 – 15 Selective Distribution

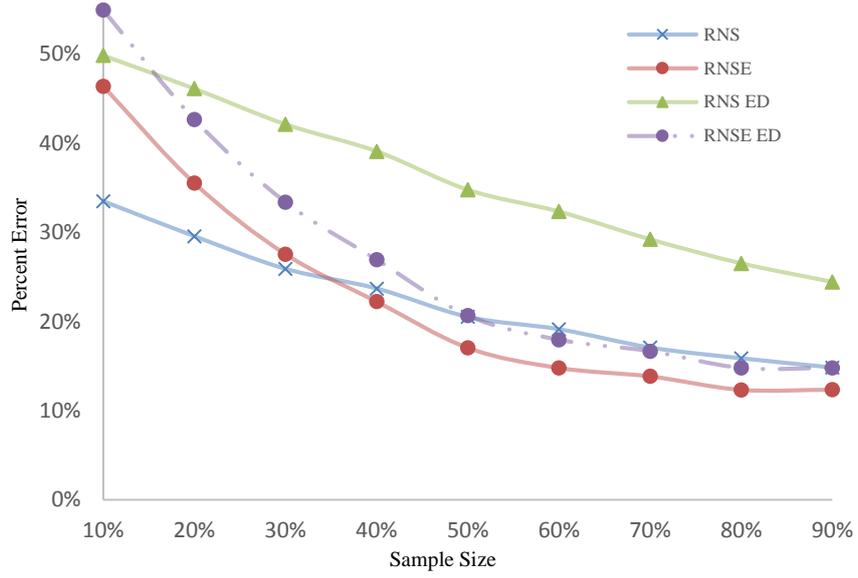

**Fig. 4.** RNS85-15 and RNSE85-15 Relative Error

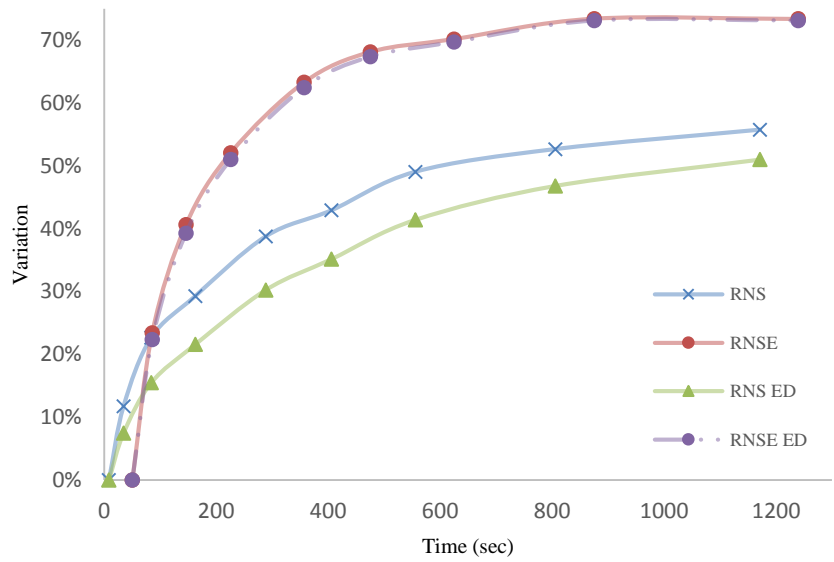

**Fig. 5.** RNS85-15 and RNSE85-15 Result Improvement over Time

### A3. 80 – 20 Selective Distribution

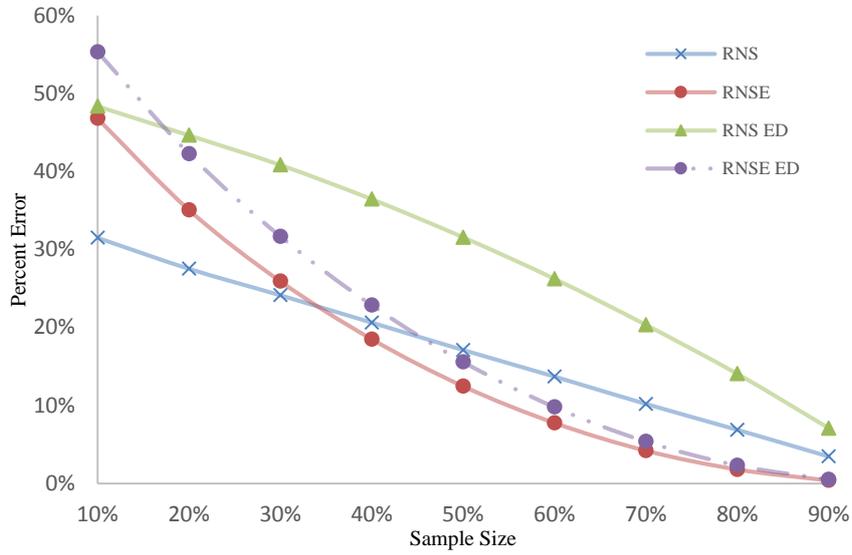

**Fig. 6.** RNS80-20 and RNSE80-20 Relative Error

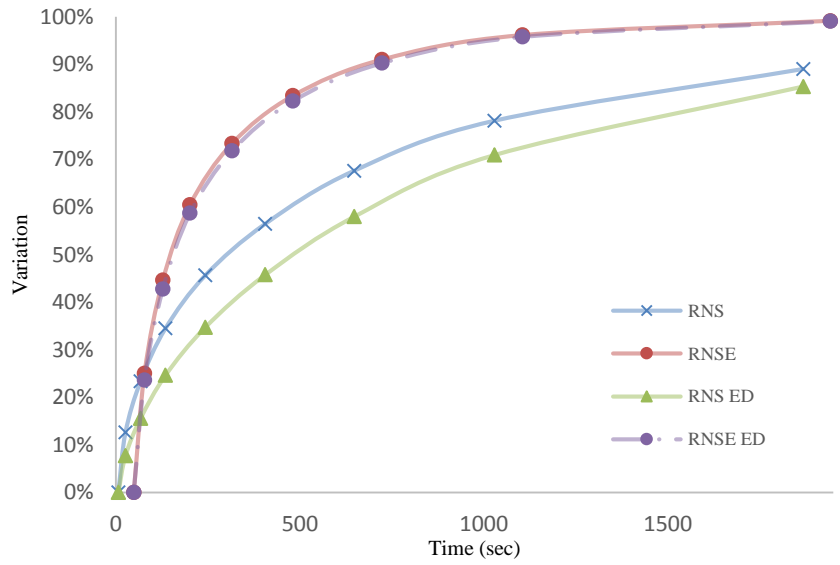

**Fig. 7.** RNS80-20 and RNSE80-20 Result Improvement over Time

### A4. 75 – 25 Selective Distribution

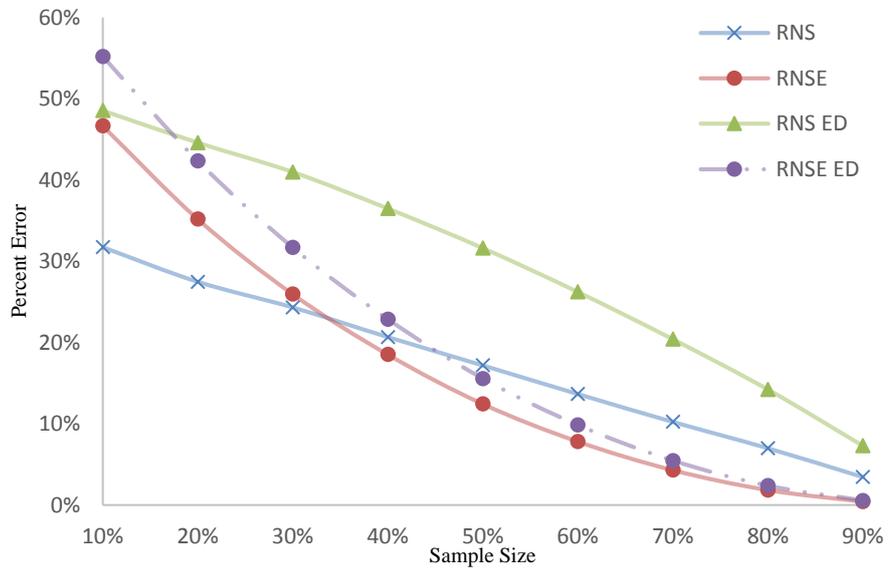

**Fig. 8.** RNS75-25 and RNSE75-25 Relative Error

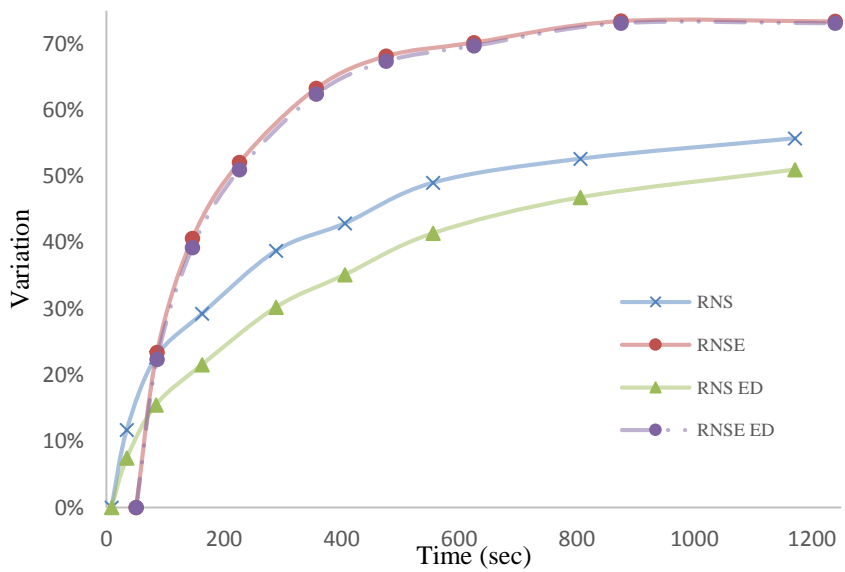

**Fig. 9.** RNS75-25 and RNSE75-25 Result Improvement over Time